\documentclass[12pt,epsfig]{iopart}
\usepackage{epsfig}
\usepackage{graphicx}
\usepackage{color}
\usepackage{mathrsfs}
\usepackage{amssymb}
\usepackage{bm}

\begin{document}

\title{\bf Quantum interference in the time-of-flight distribution}

\author{Md. Manirul Ali and Hsi-Sheng Goan}

\address{Department of Physics, Center for Quantum Science and Engineering, and Center 
for Theoretical Sciences, National Taiwan University, Taipei 10617, Taiwan}
\ead{goan@phys.ntu.edu.tw}
\begin{abstract}
We propose a scheme to experimentally observe matter-wave interference in the time domain, 
specifically in the arrival-time or the time-of-flight (TOF) distribution for atomic BEC 
Schr\"{o}dinger-cat state represented by superposition of macroscopically separated wave 
packets in space. This is in contrast to interference in space at a fixed time observed in 
reported BEC experiments. We predict and quantify the quantum interference in the TOF 
distribution calculated from the modulus of the quantum probability current density 
(rather than the TOF distributions obtained from a purely classical or semi-classical treatment 
in many reported experiments). The interference and hence the coherence in the quantum TOF signal 
disappears in the large-mass limit. Our scheme has the potential to probe the validity of various 
other theoretical approaches ({\it Phys. Rep.} {\bf 338}, 353 (2000)) of calculating the quantum 
arrival time distribution.         

PACS number(s): 03.65.Xp, 03.75.-b, 03.65.Ta

\end{abstract}

\maketitle                                                                    
             
\section{Introduction}
\begin{figure}
\centering
\epsfxsize=3.7in\epsfysize=2.7in
\epsfbox{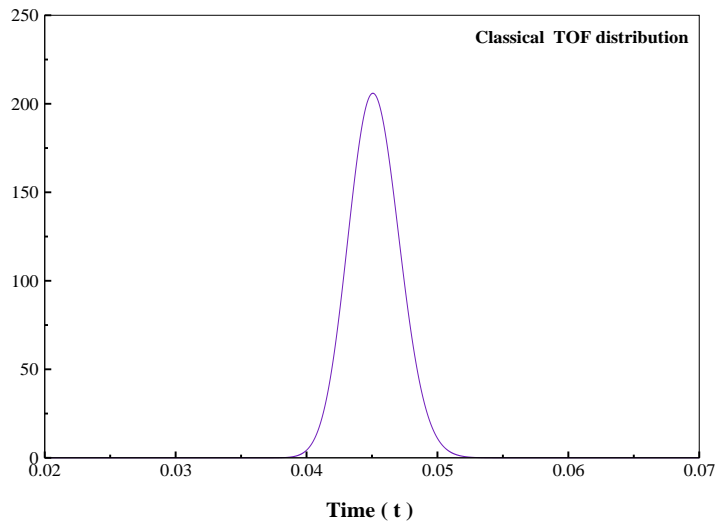}
\caption{(Color online) {Classical TOF distribution ${\mathscr D}(t)$ given by Eq.(\ref{classi}) 
is plotted for BEC of sodium atoms at a temperature $T=1~\mu K$. The detector is located at a distance
$z=H=-1$~cm and $\sigma_0=1~\mu$m. }}
\label{curvetof}
\end{figure}

In recent times, laser cooling and trapping of atoms has become an area of active research \cite{laser}. 
The temperature of the cold atomic sample is one of its most important characteristics and 
several methods have been proposed and used for its determination. A well-known technique 
of measuring this temperature is the time-of-flight (TOF) method \cite{yavin}. It is significant
to mention that the first evidence for Bose Einstein condensate (BEC) was emerged from TOF
measurements \cite{bectemp}. Most of the samples of cold atoms are initially prepared in magneto-optical 
traps and the atomic cloud is allowed for a thermal expansion after its release from the trap. These 
so-called time-of-flight (TOF) measurements are performed either by acquiring the absorption signal 
of the probe laser beam through the falling and expanding atomic cloud, or by measuring the fluorescence 
of the atoms excited by the resonant probe light. Most of the theoretical analyses of TOF measurements 
are as follows. To find the shape of the absorption TOF signal, one assumes to start with the initial Gaussian 
position and velocity distributions of atoms in the trapped sample. The initial probability distribution 
of finding an atom in the phase space volume element with coordinates ($z_0$, $v_{0}$) is given by
\begin{eqnarray}
\nonumber
{\mathscr D}(z_0,v_0) d z_0 d v_0 &=& \frac{1}{(2 \pi {\sigma}_0^2)^{1/2}}
\exp\left(-\frac{z_0^2}{2{\sigma}_0^2}\right)\\
&\times &\frac{1}{(2 \pi {\sigma}_v^2)^{1/2}} 
\exp\left(-\frac{v_0^2}{2{\sigma}_v^2}\right) d z_0 d v_0 
\label{tofdist1}
\end{eqnarray}
\noindent
Here for simplicity, we consider the one-dimensional case.
The Gaussian width $\sigma_v$ of the velocity distribution is associated with the temperature
$T$ of the cloud by the relation ${\sigma}_v^2={kT}/{m}$, where $m$ stands for the atomic mass 
and $k$ is the Boltzmann constant. Using the Newton's equations for ballistic motion of a particle 
accelerated by the earth's gravitational field (in the vertical $z$-direction), the velocity is 
obtained in terms of the time of flight as
\begin{equation}
v_0=(z- z_0 +\frac{1}{2}g t^2)/t~~, ~~~ \frac{\partial v_0}{\partial t}= 
\frac{(z_0 + \frac{1}{2}g t^2 - z)}{t^2}.
\label{tofdist2}
\end{equation}
\noindent
Substituting the above expression for $v_0$ from Eq.(\ref{tofdist2}) in Eq.(\ref{tofdist1}), and 
then finally integrating over $z_0$, one can obtain the TOF distribution at an arbitrary 
distance $z=H$, given by
\begin{eqnarray}
\nonumber
{\mathscr D}(t) d t &=& \frac{1}{(2 \pi ~ t^2)^{1/2}} \frac{\left(\frac{1}{2}g t^2(2{\sigma}_0^2 
+ \sigma_v^2 ~ t^2) - H \sigma_v^2 ~ t^2 \right)}{\left({\sigma}_0^2 + \sigma_v^2 ~ t^2 \right)^{3/2}}\\
&\times& \exp \left(-\frac{(H+\frac{1}{2}g t^2)^2}{2({\sigma}_0^2 + \sigma_v^2 ~ t^2)} \right) d t.
\label{classi} 
\end{eqnarray}
\noindent
Figure \ref{curvetof} shows a typical TOF or arrival time distribution for cold sodium atoms.
This kind of purely classical analyses are adopted in most of the discussions on TOF
measurements where arrival time of atomic or sub-atomic particles is treated as an elementary
well-defined, unique, and classical quantity. Also, the theoretical treatments of 
the TOF distribution that can be obtained using, for instance, the Green's function method \cite{weiss} 
or any semiclassical method \cite{viana}, however, are equivalent to the TOF distribution obtained 
by using Newton's equations for ballistic motion of particles \cite{yavin}. The interpretations or 
theoretical analyses of the results of the various TOF experiments \cite{iondna,temperature,tof} with 
molecular, atomic or sub-atomic particles where classical trajectories are inferred from Newtonian 
mechanics remain debatable, especially in the domain of small atomic masses and low temperatures 
where quantum mechanical effects should be significant and quantum TOF distribution can not be 
reproduced with classical or semi-classical analyses.

\vskip 0.5cm
\noindent
Here we provide an example in the context of BEC matter-wave interference, where a quantum analysis 
for TOF is necessary. We propose a scheme for measuring the TOF distribution of a freely falling
atomic BEC prepared in non-classical Schr\"{o}dinger-cat states. The interference in the TOF 
distribution (or signal) can then be observed by taking a note or record of the particle counts over 
various {\it tiny time-windows} at a fixed detector location. 
This is different from the interference between two freely expanding 
BEC's observed \cite{andrews} in {\it space} after a definite 
time of free fall of the condensates. Coherent splitting of BEC atoms with optically induced Bragg 
diffraction have been done experimentally \cite{kozuma,prl}. 
The {\it spatial coherence} of a BEC is measured 
using interference technique by creating and recombining two spatially displaced, coherently diffracted 
copies of an original BEC \cite{prl}. 

\vskip 0.5cm
\noindent
As mentioned above, most of the experiments  (particularly when matter-waves are associated with centre-of-mass 
motion or external motion of massive quantum particles) demonstrate matter-wave interference by showing the intensity 
variation at an extended region of detection space at a fixed time. In the present paper, in contrast, we predict and 
quantify the matter-wave interference in the center-of-mass motion by calculating the {\it time distribution} of 
matter-wave arrival probability at some fixed spatial point. More specifically, we discuss here the BEC matter-wave 
interference in the TOF distribution or {\it arrival time distribution} since BEC as a source of coherent matter waves 
is already routinely demonstrated and thus may be an ideal candidate to show interference signal in the time domain 
(arrival-time distribution). We use here a particular quantum approach to calculate the TOF distribution and in our analysis 
we do not use at any point classical or semiclassical ingredients. We consider the free fall of matter-wave associated 
to quantum particles represented by an initial Schr\"{o}dinger-cat state which is the linear superposition of two 
mesoscopically distinguishable Gaussian wave packets peaked around different heights viz., $z=0$ and $z=-d$ along the 
vertical z-axis. Then after a certain height of free fall (evolution under the potential $V=mgz$) of the Schr\"{o}dinger 
cat, we calculate the quantum TOF distribution at a given detector location $z=H$. During the free fall, the distinct 
superposed wave packets of the Schr\"{o}dinger cat overlap or interfere in space, so it is natural to expect that they 
will also interfere in the {\it time of fall} showing an interference pattern in the quantum TOF distribution. 

\vskip 0.5cm
\noindent
We take this particular example of matter-wave interference in the discussion of quantum TOF distribution to pinpoint 
the necessity of a quantum analysis. So, the need for a quantum analysis of TOF distribution is not merely a conceptual 
but a practical issue, asking how to predict the TOF distribution using only classical and semi-classical ingredients 
in a purely quantum scenario like this (interference in the TOF distribution for quantum particles). Now, in spite of 
the emphasis of quantum theory on the observable concept, there is no commonly accepted recipe to incorporate time 
observables and their probability distributions in the quantum formalism, and there is considerable difficulty and debate 
over the issue of defining time (for example, tunneling time, decay time, arrival time) as an observable 
\cite{timebook,tunnel,kijowski,operator,timemeas,timereport,current,rotator,finkel,bohmarrtim,ali1,ali2,class,weq}. 
Even for the simplest case of arrival time problem there is no unique way to calculate 
the probability distribution in the quantum formalism \cite{timereport}. Despite this, many researchers have evidently 
not been discouraged from seeking an expression for the arrival time distribution (or the quantum TOF distribution) 
within a consistent theoretical framework. Several logically consistent schemes for the treatment of the arrival 
time distribution have been formulated, such as those based on axiomatic approaches \cite{kijowski}, 
operator constructions \cite{operator}, measurement based approaches \cite{timemeas,rotator}, trajectory 
models \cite{bohmarrtim} and probability current density approach 
\cite{timereport,current,rotator,finkel,bohmarrtim,ali1,ali2,class,weq}. We will 
use here probability current density approach to calculate the quantum TOF distribution which is logically 
consistent and also physically motivating.

\vskip 0.5cm
\noindent
The main purpose of our paper is two-fold. First, our proposal to experimentally observe or quantitatively predict 
the matter-wave interference in the time domain, specifically in the TOF distribution is itself quite significant 
which has not been explored in the current literature to the best of our knowledge. BEC as a source of coherent 
matter waves is already routinely demonstrated in spatial interference experiments, so BEC will also be an ideal 
candidate to show the interference in the TOF distribution. Second, we have just mentioned that there is an inherent 
nonuniqueness within the formalism of quantum mechanics for calculating the TOF or {\it arrival time} distribution. 
It remains an open question as to what extent these different quantum mechanical approaches 
\cite{timebook,tunnel,kijowski,operator,timemeas,timereport,current,rotator,finkel,bohmarrtim,ali1,ali2,class,weq}
for calculating the time distributions can be tested or empirically discriminated. Our proposal of measuring matter-wave 
interference in TOF distribution has the potential to empirically resolve ambiguities inherent in the theoretical 
formulations of the quantum TOF distribution. In this respect, it would be interesting if the prediction of BEC matter 
wave {\it interference} in the TOF distribution (calculated from different quantum approaches) be verified in actual experiments.

\section{Interference in the quantum time-of-flight distribution for Bose-Einstein Condensate}
\label{sec:InterferenceTOF}

We begin our analysis with the standard description of the flow of probability in quantum mechanics, 
which is governed by the continuity equation derived from the Schr\"{o}dinger equation given by

\begin{equation}
\frac{\partial}{\partial t}|\Psi({\bf x},t)|^2 + {\bf \nabla}.{\bf J}({\bf x},t)=0
\label{continuity}
\end{equation}

\noindent
The quantity ${\bf J}({\bf x},t)$=$\frac{i\hbar}{2m}(\Psi {\bf \nabla}\Psi^{\ast}-\Psi^{\ast}{\bf \nabla} \Psi)$ 
defined as the probability current density corresponds to this flow of probability. In one dimension, the current 
density $J(x,t)$ tells us the rate at which probability is flowing past the point $x$. So, interpreting the one 
dimensional continuity equation in terms of the flow of physical probability, the Born interpretation for the 
squared modulus of the wave function and its time derivative suggest that the arrival time distribution of the 
particles reaching a detector located at $x=X$ can be calculated 
\cite{timereport,current,rotator,finkel,bohmarrtim,ali1,ali2,class,weq} using the probability 
current density $J(x,t)$. It should also be noted that $J(x,t)$ can be negative, hence one needs to take the modulus 
sign in order to use the above definition. Our aim here is to derive an expression for the TOF distribution through 
the quantum probability current density for the atomic BEC representing the mesoscopic Schr\"{o}dinger cat and 
showing the interference in TOF signal. Probability current density approach to the TOF distribution is also 
justified by the Bohmian model of quantum mechanics in terms of the causal trajectories of individual particles 
\cite{bohmarrtim}. Although the Schr\"{o}dinger probability current density is formally nonunique up 
to a total divergence term \cite{finkel}, the current can be uniquely fixed if one calculate the current 
in the non-relativistic limit of a proper relativistic wave equation which provide appropriate spin-dependent 
corrections to it \cite{ali2,holland}. We ignore this small spin-dependent contribution here in our present 
discussion, as the estimated magnitude of the spin-dependent current is roughly $10^{5}$ to $10^{6}$ times 
smaller than the Schr\"{o}dinger current. It was emphasized that the probability current density approach not only 
provides an unambiguous definition of arrival time at the quantum mechanical level 
\cite{timereport,current,rotator,finkel,bohmarrtim,ali1,ali2,class,weq}, but also adresses the issue of obtaining 
the proper classical limit of the TOF of massive quantum particles \cite{class,weq}.

\vskip 0.5cm
\noindent
Now, to keep the discussion concise, here we restrict ourselves to the case of one-dimensional motion, but our 
resulting conclusion does not depend on three-dimensional extension which will be straightforward as discussed 
later at the end of this Section. A magnetically trapped BEC as a source of coherent matter wave or atom laser, 
where a macroscopic number of atoms occupy the same ground state is now routinely available. After being released 
from the trapped Bose-Einstein condensates, matter waves fall freely due to the gravity. If the atomic beam is well 
collimated, we can use a one-dimensional Gross-Pitaevskii equation \cite{bectemp,japan} for the evolution of 
condensate wavefunction $\Psi(z,t)$ with the gravitational potential,
\begin{eqnarray}
\fl i \hbar \frac{\partial}{\partial t} \Psi(z,t) = - \frac{\hbar^2}{2m} \frac{\partial^2}{\partial z^2} \Psi(z,t)
+  mgz \Psi(z,t) + U_0 |\Psi(z,t)|^2 ~\Psi(z,t),
\label{pita}
\end{eqnarray}
where $|\Psi(z,t)|^2$ provides the density profile of the BEC, $m$ denotes the atomic mass, $g$ the gravity 
acceleration, and $U_0$ the inter-atomic interaction strength. In our present discussion we consider condensate 
of non-interacting bosons and we neglect \cite{japan,nonint} the effects of inter-atomic interaction $U_0$ on 
the freely falling condensate. In the BEC, the whole complex is described by one single wave function $\Psi(z,t)$ 
(a macroscopic wave function of the condensate) exactly as in a single atom, and we can speak of coherent matter 
in the same way as of coherent light in the case of a laser. To show {\it interference} in the quantum TOF signal 
for the freely falling BEC, we consider the initial state of the BEC be prepared in a Schr\"{o}dinger-cat state 
which is the coherent superposition of two mesoscopically distinguishable states in the configuration space: 
\begin{eqnarray}
\Psi(z,0)={\mathscr N} \left[c_1 \psi_1(z,0)+ c_2 \psi_2(z,0) \right],
\label{wavefn0}
\end{eqnarray}
\noindent
where
\begin{eqnarray}
\psi_1(z,0)=\frac{1}{(2 \pi {\sigma}_0^2)^{1/4}} \exp\left(-\frac{z^2}
{4{\sigma_0}^2}\right),
\end{eqnarray}
\begin{eqnarray}
\psi_2(z,0)=\frac{1}{(2 \pi {\sigma}_0^2)^{1/4}} \exp\left(-\frac{(z+d)^2}
{4{\sigma_0}^2}\right)
\end{eqnarray}
\noindent
are the Gaussian wave packets centered around $z=0$ and $z=-d$, respectively, and $\sigma_0$ is the initial position 
spread. A description of the initial 1D wavefunction for two separated BEC's using the Gaussian form (\ref{wavefn0}) 
has been made, for example, by the authors of \cite{gaussian}, where they consider all the non-interacting bosons are 
prepared to be condensed in the ground state of the harmonic trap \cite{bectemp}. For simplicity, we take $c_1=c_2
=1/\sqrt{2}$ which implies that after coherent splitting of the original BEC, each component has equal number of atoms. 
Then the value of the normalization constant 
\begin{equation}
{\mathscr N}=1/\sqrt{1+\exp(-d^2/8{\sigma_0}^2)}.
\label{normal}
\end{equation}
\noindent
As we have mentioned, the Schr\"{o}dinger-cat state of matter was generated for a BEC represented by superposition 
of spatially separated states and the superposition was verified \cite{prl} by detecting the quantum mechanical 
interference ({\it in space}) between the localized wave packets separated by a mesoscopic distance. Under the 
experimental situtations (where the spatial coherence of the BEC was measured using interference technique by creating 
and recombining two spatially displaced, coherently diffracted copies of an original BEC) discussed by the authors of 
\cite{kozuma,prl}, the BEC wave function can be written as a linear superposition of spatially separated wave packets  
\cite{gaussian} which may be inferred as a true macroscopic Schr\"{o}dinger cat. Now considering the free fall of the 
coherently splitted BEC under gravity, we calculate the time evolution of the Schr\"{o}dinger-cat state (\ref{wavefn0}) 
according to equation (\ref{pita}) with $U_0=0$, we then obtain
\begin{eqnarray}
\Psi(z,t)=\frac{{\mathscr N}}{\sqrt{2}} \left[ \psi_1(z,t) + \psi_2(z,t) \right],
\label{wavefn1}
\end{eqnarray}
where
\begin{eqnarray}
\fl \psi_1(z,t) = \left(2\pi s^{2}_{t}\right)^{-1/4}
\exp \left[ \frac{ -\left( z + \frac{1}{2}g t^2\right)^2}{4s_{t}\sigma_{0}} \right] 
\exp \left[-i(\frac{m}{\hbar}) \left(gtz + \frac{1}{6}g^2 t^3 \right) \right],
\end{eqnarray}
and
\begin{eqnarray}
\fl \psi_2(z,t) = \left(2\pi s^{2}_{t}\right)^{-1/4}
\exp \left[ \frac{ -\left( z + d + \frac{1}{2}g t^2\right)^2}{4s_{t}\sigma_{0}} \right] 
\exp \left[-i(\frac{m}{\hbar}) \left(gtz + \frac{1}{6}g^2 t^3 \right) \right]
\end{eqnarray}
with
\begin{eqnarray}
s_{t}=\sigma_{0}\left(1+i\hbar t/2{m}\sigma_{0}^{2}\right).
\label{st}
\end{eqnarray}
The expression for the Schr\"{o}dinger probability current density corresponding to the time evolved 
state $\Psi(z,t)$ (\ref{wavefn1}) is given by
\begin{eqnarray}
\fl J(z,t) = \frac{i\hbar}{2m}(\Psi \frac{\partial \Psi^{\ast}}{\partial z}
-\Psi^{\ast}\frac{\partial \Psi}{\partial z})
= \frac{{\mathscr N}^2}{2} \left[ J_1(z,t)+J_2(z,t)+J_3(z,t)+{J_3}^{\ast}(z,t) \right],
\label{pure}
\end{eqnarray}
where 
\begin{eqnarray}
J_1(z,t) = \left[ \frac{{\hbar}^2 t}{4 m^2 {\sigma_0}^2 {\sigma}^2}( z +
\frac{1}{2}g t^2) -gt \right] \times |\psi_1(z,t)|^2, 
\end{eqnarray}
\begin{eqnarray}
J_2(z,t) = \left[ \frac{{\hbar}^2 t}{4 m^2 {\sigma_0}^2 {\sigma}^2}( z +
d + \frac{1}{2}g t^2) -gt \right] \times |\psi_2(z,t)|^2, 
\end{eqnarray}
\begin{eqnarray}
J_3(z,t)+{J_3}^{\ast}(z,t) = 2~ P_{12}(z,t) \left(\eta \cos\delta - \lambda \sin\delta \right),
\label{purecross}
\end{eqnarray}
where
\begin{eqnarray}
\lambda=\frac{\hbar d}{4 m {\sigma}^2},
\label{lambda1}
\end{eqnarray}
\begin{eqnarray}
\eta=\frac{{\hbar}^2 t}{8 m^2{\sigma_0}^2 {\sigma}^2} \left(2z + d +g t^2  \right) -gt,
\label{eta1}
\end{eqnarray}
and
\begin{eqnarray}
P_{12}(z,t)=|\psi_1(z,t)|~|\psi_2(z,t)|
\label{overlap1}
\end{eqnarray}
with the time-dependent position spread given by
\begin{eqnarray}
\sigma^2=s_t s_{t}^{\ast}=\sigma^2_{0}\left(1+\hbar^{2}t^{2}/4{m}^{2}\sigma_{0}^{4}\right).
\label{spread}
\end{eqnarray}
Here the quantity $s_t$ is defined in equation (\ref{st}).
The oscillatory factor $\delta$ in (\ref{purecross}), responsible for the interference
effect, is given by
\begin{eqnarray}
\delta=\frac{\hbar t}{8 m {\sigma_0}^2 {\sigma}^2}\left(d^2 + dg t^2 + 2zd  \right)~
=~\frac{\hbar t \left( d^2 + d g t^2 + 2 z d \right)}{8 m \left(\sigma^4_0 + \frac{\hbar^2 t^2}{4 m^2}\right)}.
\label{delta}
\end{eqnarray}
Now taking the modulus of the quantum probability current density (\ref{pure}), we obtain the quantum 
TOF distribution \cite{timereport,current,rotator,finkel,bohmarrtim,ali1,ali2,class,weq} at a detector 
location $z=H$ for the spatially separated BEC Schr\"{o}dinger cat falling freely under gravity given by
\begin{eqnarray}
{\Pi}(t)=|J(z=H,t)|.
\label{pi}
\end{eqnarray}

\vskip 0.5cm
\noindent
Exactly similar expression for quantum TOF distribution can be obtained for a three dimensional analysis of the 
problem by considering the form of the initial wave function 
\begin{eqnarray}
\Psi(x,y,z,0)=\frac{{\mathscr N}}{\sqrt{2}} \left[ \psi_1(x,y,z,0) + \psi_2(x,y,z,0) \right],
\end{eqnarray}
\begin{eqnarray}
\fl \psi_1(x,y,z,0)=\frac{1}{(2 \pi {\sigma}_0^2)^{3/4}} \exp\left(-\frac{x^2}{4{\sigma_0}^2}\right)
\exp\left(-\frac{y^2}{4{\sigma_0}^2}\right) \exp\left(-\frac{z^2}{4{\sigma_0}^2}\right),
\end{eqnarray}
\begin{eqnarray}
\fl \psi_2(x,y,z,0)=\frac{1}{(2 \pi {\sigma}_0^2)^{3/4}} \exp\left(-\frac{x^2}{4{\sigma_0}^2}\right)
\exp\left(-\frac{y^2}{4{\sigma_0}^2}\right) \exp\left(-\frac{(z+d)^2}{4{\sigma_0}^2}\right),
\end{eqnarray}
where $\psi_1(x,y,z,0)$ and $\psi_2(x,y,z,0)$ are now three dimensional Gaussian wave packets
separated along the vertical ${\bf \widehat z}$-axis, having peaks around the points $(0,0,0)$ and 
$(0,0,-d)$ respectively. The value of the normalization constant ${\mathscr N}$ remains the same as 
that in Eq.(\ref{normal}). One can then obtain the three dimensional Schr\"{o}dinger time evolved wave 
function under the gravitational potential. The quantum TOF distribution can then be calculated again 
using the three dimensional quantum current. Interpreting again the three dimensional continuity equation 
(\ref{continuity}) in terms of the flow of physical probability, one can define the quantum TOF 
distribution for the atoms crossing a surface element $d{\bf S}$ as $|{\bf J}.d{\bf S}|$. It is important 
to mention here that the quantum flux density $|{\bf J}.d{\bf S}|$ have been identified with the 
``time distribution'' of particles crossing the surface element $d{\bf S}$ by Daumer {\it et al.} \cite{daumer}, 
who were applying Bohm model to the scattering problem for a quantum particle in three dimension. Hence quantum 
TOF distribution for the atoms reaching at a finite surface plane $S$ in three-dimension will be given by
\begin{eqnarray}
{\bm \Pi}(t) = |~ \int_S \int {\bf J}.d{\bf S}~~| = |~ \int_S \int {\bf J}.{\bf \widehat n}~dS~~|,
\label{tof}
\end{eqnarray}
\noindent
where ${\bf \widehat n}$ is the unit vector normal to the surface. The quantum TOF distribution for the atoms 
reaching the XY plane (${\bf \widehat n}=-{\bf \widehat z}$) after a certain height ($z=H$) of free fall is then 
given by
\begin{eqnarray}
{\bm \Pi}_1 (t)=|~ \int_S \int {\bf J}.{\bf \widehat n}~dS~~| = |~\int_{x} \int_{y} J_{z}(x,y,z=H,t) dx dy~~|, 
\label{pi1}
\end{eqnarray}
\noindent
where $J_{z}(x,y,z=H,t)$ is the z-component of the three dimensional probability current density at a fixed height $z=H$. By 
evaluating the integral of Eq.(\ref{pi1}), one can see that ${\bm \Pi}_1 (t)$ is exactly the same as the quantum 
TOF distribution ${\Pi}(t)$ obtained for one dimensional analysis (\ref{pi}). To understant more clearly the origin
of this interference in TOF, let us consider the propagation (evolution) of individual wave packets $\psi_1$ and 
$\psi_2$ under the gravitational potential. Then one will have two distinct TOF distributions having separate mean 
arrival times. This is because the peaks of the component wave packets $\psi_1$ and $\psi_2$ take different times 
to reach the detector at $z=H$, since they are spatially separated along the vertical z-axis, and the interference 
in TOF arises due to the superposition of these two wave packets. In this setup, the cross term (interfering term) 
in the quantum TOF distribution arises from the relative phase of the component wave packets ($\psi_1$ and $\psi_2$) 
along z-direction, as only z-componets of the component wave packets differ in the time evolution and continue to 
develope the relative phase, and this relative phase is not cancelled out when we perform the integration over XY-plane.
The interference pattern in the quantum TOF distribution can be detected by using a probe laser, focused in the form 
of a sheet underneath the falling BEC atoms in the XY-plane at $z=H$. When the trapping forces are turned off, the 
BEC atoms will fall through the laser probe under the influence of gravity. It is then possible to detect the fluorescence 
from the atoms excited by the resonant probe light as they reach the detection sheet. The fluorescence can be measured 
as a function of time to determine the TOF distribution. For this setup, one can also consider a situation where the 
detection is made in the YZ-plane (${\bf \widehat n}=-{\bf \widehat x}$) at a fixed $x=X$. In that case, quantum TOF 
distribution (say, ${\bm \Pi}_2(t)$) can be obtained from the x-component of the three dimensional current 
($J_{x}(x=X,y,z,t)$) integrated over YZ-plane using Eq.(\ref{tof}). By evaluating that integral, one can see that 
there will be no interference in the quantum TOF distribution ${\bm \Pi}_2(t)$ under this situation. This is because 
the interference term in the quantum TOF distribution is wiped out when we perform the integration over the YZ-plane. 

\vskip 0.5cm
\noindent
Next, we consider another setup in three dimension where the superposed wave packets $\psi_1(x,y,z,0)$ and
$\psi_2(x,y,z,0)$ are separated along the horizontal X-axis, having peaks around the points $(0,0,0)$ and
$(-d,0,0)$ respectively. This situation is analogous to the experimental setup of Ref.\cite{andrews} where
BEC interference was observed in {\it space} at a fixed time. For this geometry, one can again calculate the 
three dimensional Schr\"{o}dinger time evolved wave function under the gravitational potential. The quantum 
TOF distribution (\ref{tof}) can then be calculated again using the three dimensional quantum current. For 
this setup, we again consider the detection of the particles at a surface plane (XY-plane with ${\bf \widehat n}
=-{\bf \widehat z}$) at $z=H$. In this case, quantum TOF distribution (say, ${\bm \Pi}_3(t)$) can be obtained 
from the z-component of the three dimensional current integrated over XY-plane using Eq.~(\ref{tof}). By
evaluating that integral, one can see that there will not be any interference at all in the quantum TOF 
distribution ${\bm \Pi}_3(t)$. The interference term in the quantum TOF distribution is wiped out when we perform 
the integration over the XY-plane. This is because for this setup, the individual wave packets $\psi_1$ and $\psi_2$ are not 
separated along the vertical z-axis, so they will have 
the same TOF distribution 
with the same mean arrival time to reach the detection plane at $z=H$. 
Hence, we do not expect any interference in the quantum TOF distribution detected at the horizotal plane at $z=H$ when we consider 
the superposition of the horizontally separated wave packets.
The interference term in the quantum TOF distribution, in this case, 
is wiped out when we perform 
the integration over the XY-plane, even though one can observe the interference in space at a fixed time. For this 
setup, one can also consider a situation where the detection is made in the YZ-plane (${\bf \widehat n}
=-{\bf \widehat x}$) at a fixed $x=X$. In that case, quantum TOF distribution (say, ${\bm \Pi}_4(t)$) can be 
obtained from the x-component of the three dimensional current integrated over YZ-plane using Eq.(\ref{tof}). 
For this situation, although we will see the presence of some interfering terms in the expression of quantum 
TOF distribution ${\bm \Pi}_4(t)$, the intensity to observe this interference will be very low, as only a 
small fraction of the condensate atoms will arrive at the detection YZ-plane at $x=X$ due to free expansion 
(free particle motion) of the wave packets. One can also check that the quantum TOF distribution ${\bm \Pi}_4(t)$ 
will have exactly the same expression as $\Pi(t)$ of (\ref{pi}) with $g=0$ (no gravity) and with $z=H$ replaced 
by $x=X$.
We will show numerically in the next section that 
gravity plays an important role in our setup
to pull down the vertically separated superposed condensate 
toward the detection plane at $z=H$.

\vskip 0.5cm
\noindent
Hence the only two situations (in our above discussion) where we see the presence of interference 
in three dimensions are the quantum TOF distribution ${\bm \Pi}_1(t)$ and ${\bm \Pi}_4(t)$. Now, 
${\bm \Pi}_1(t)$ is exactly the same as $\Pi(t)$ and ${\bm \Pi}_4(t)$ is also same as $\Pi(t)$ with $g=0$ 
and with $z=H$ replaced by $x=X$. So the whole characteristic of the interference pattern in quantum TOF 
distribution hinges upon the form of $\Pi(t)$. In the next section we study numerically the parameter 
dependence of the quantum TOF distribution $\Pi(t)$ and the physical interplay between these parameters.

\section{Numerical Results and Discussions}

\begin{figure*}
\includegraphics{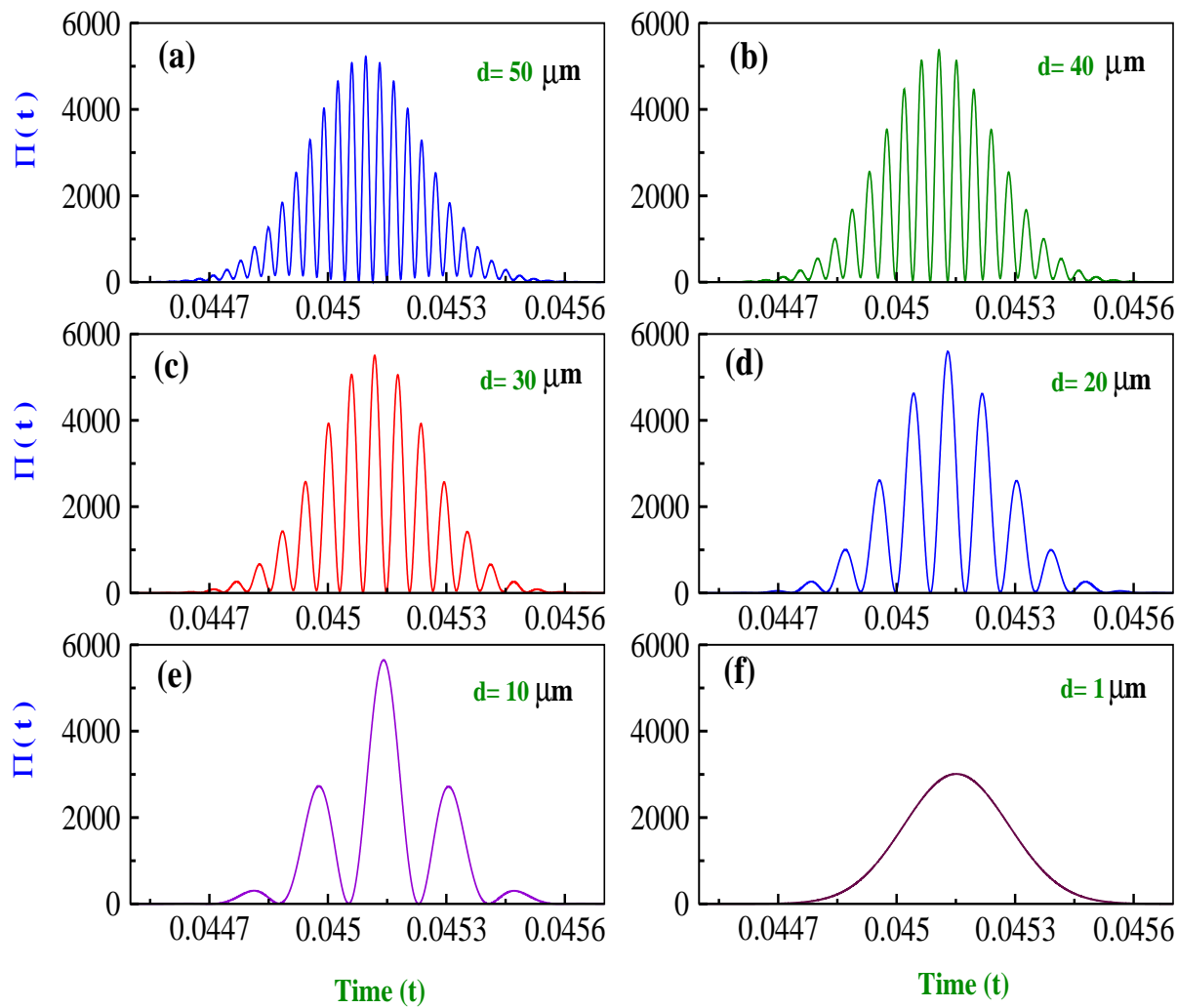}
\caption{(Color online) Quantum TOF distributions $\Pi(t)$ for the coherently splitted BEC of
sodium atoms (representing the macroscopic Schr\"{o}dinger-cat) falling freely under gravity
are plotted for varying wave packet separation $d$. In each curve, time (in $sec$) is plotted along
horizontal direction and the coherent TOF distribution $\Pi(t)$ of BEC
Schr\"{o}dinger-cat is plotted (in ${sec}^{-1}$) along vertical axis. The detector is located at a
distance $z=H=-1$~cm and $\sigma_0=1~\mu$m.}
\label{BECdf}
\end{figure*}

Quantum TOF distribution ${\Pi}(t)$ of the freely falling atomic BEC Schr\"{o}dinger cat is plotted 
(Fig.\ref{BECdf}) at a detector location $z=H=-1$ cm with different values of wave packet separation 
$d$. We see clear signature of interference in the quantum TOF distribution arising due to the terms 
$J_3(z=H,t)$ and ${J_3}^{\ast}(z=H,t)$ of (\ref{purecross}) in the expression for quantum probability 
current density (\ref{pi}) and (\ref{pure}). During free fall, the spatially separated wave packets 
of the BEC Schr\"{o}dinger cat overlap or interfere in space and hence they also interfere in the 
time of fall showing an interference pattern in the quantum TOF distribution. The quantum TOF 
distribution ${\Pi}(t)$ may be visualized as a coherent pulse of BEC atoms. The interference pattern 
in the quantum TOF signal (Fig.\ref{BECdf}a) is very sharp for a typical set of parameter values, 
for example, $H=-1$ cm, $d=50~\mu$m, $\sigma_0=1~\mu$m, and the pattern disappears (Fig.\ref{BECdf}f) 
when the separation between the BEC superposed wave packets is decreased to $d=1\mu$m for the above 
mentioned parameter values. We can see from the oscillatory factor $\delta$ (\ref{delta}) in 
(\ref{purecross}) that the number of oscillations and hence the number of fringe increases in the 
TOF distribution $\Pi(t)$ (\ref{pi}) as one increase the separation $d$. 
The interference effect arises mainly because of two factors: one is the temporal 
overlap $P_{12}(z=H,t)$ (\ref{overlap1}) and the other is the oscillatory factor $\delta$. 
When $d$ is very small, the overlap $P_{12}(z=H,t)$ is very high, but the oscillatory factor $\delta$ 
becomes small. As a consequence, the oscillation frequency is too slow or the oscillation period is too 
large, and we do not see any oscillatory effect in the temporal overlap region of the wave packets. Number 
of oscillations increases as one increases $d$, but again after a certain value ($d > 400 \mu$m) of 
separation there will be no interference as the overlap $P_{12}(z=H,t)$ becomes very small in that case. 

\begin{figure*}
\includegraphics{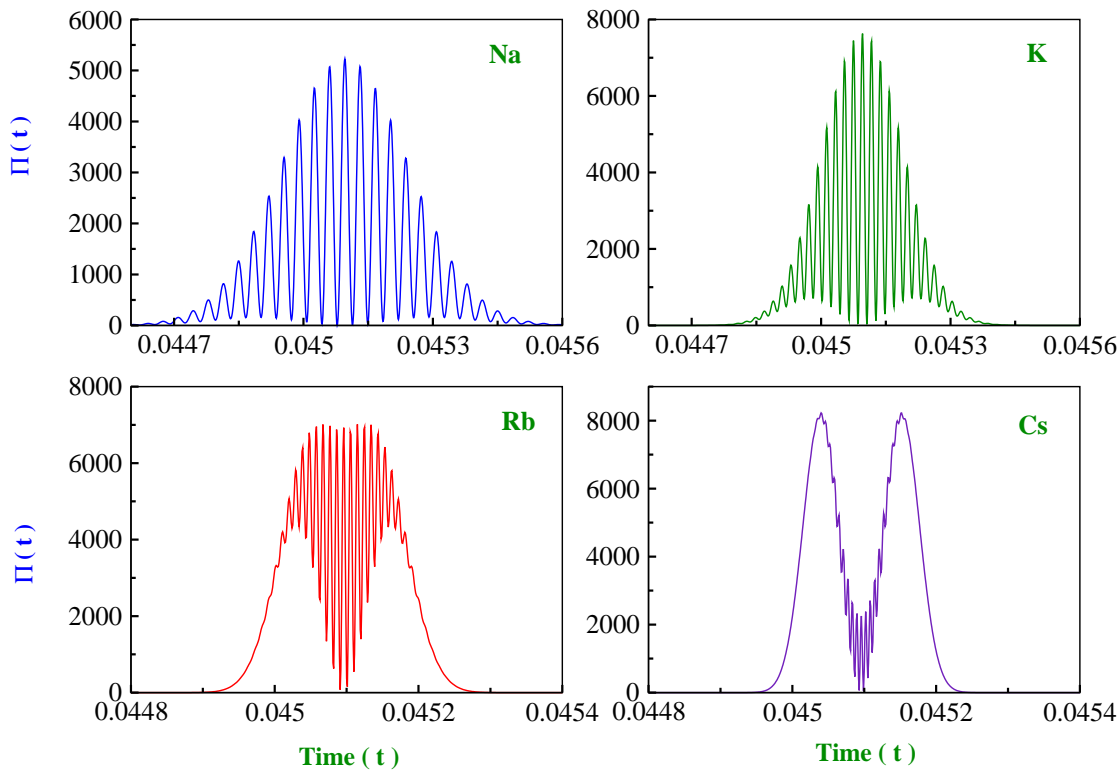}
\caption{ (Color online) Quantum TOF distributions $\Pi(t)$ for the coherently splitted BEC of
sodium atoms (representing the macroscopic Schr\"{o}dinger-cat) falling freely under gravity
are plotted for varying atomic masses. In each curve time (in $sec$) is plotted along horizontal
direction and the TOF signal $\Pi(t)$ is plotted (in ${sec}^{-1}$) along vertical axis. The detector
is located at $z=H=-1$ cm with $d=50~\mu$m and $\sigma_0=1~\mu$m.}
\label{Mass1}
\end{figure*}

\begin{figure*}
\includegraphics{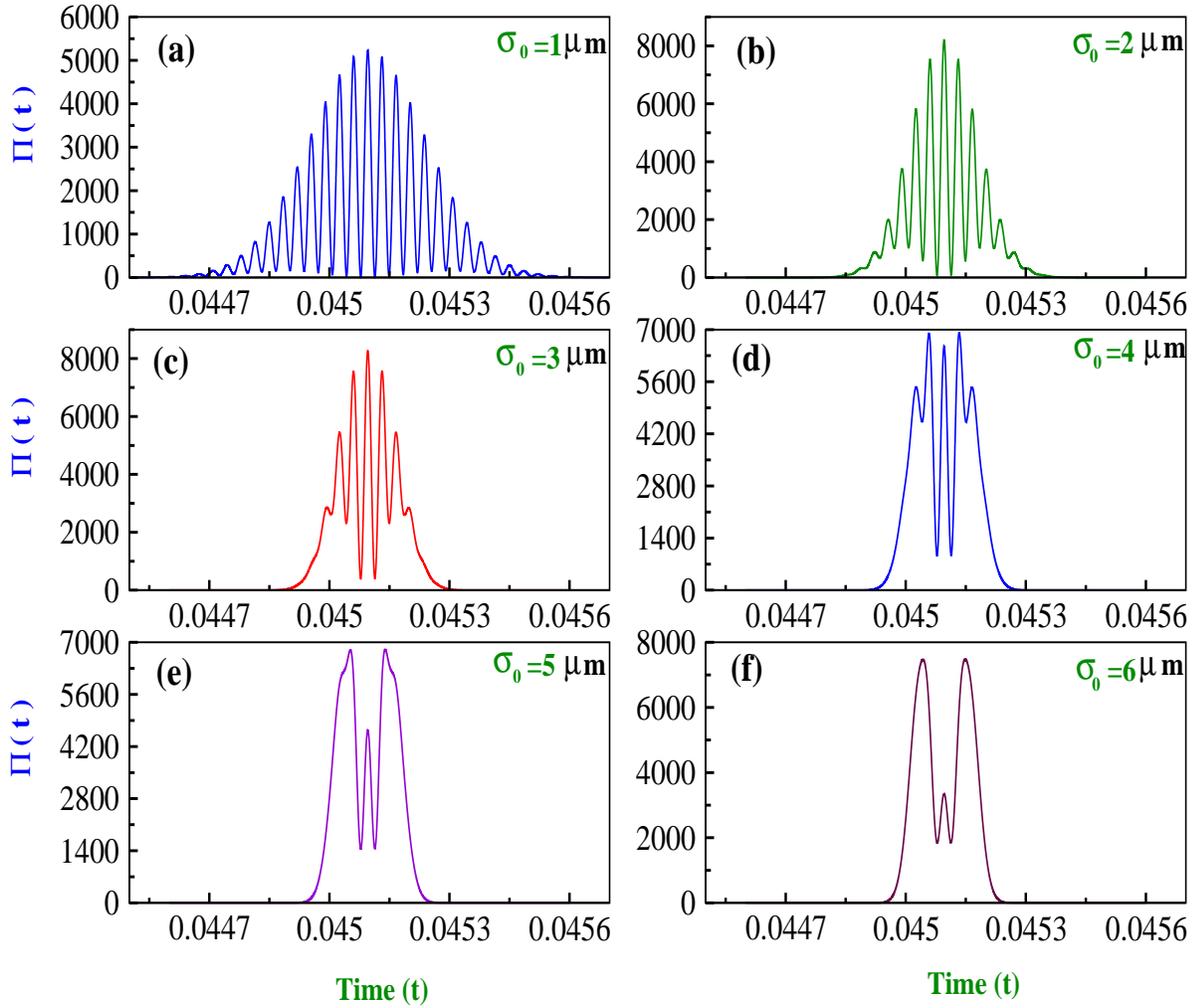}
\caption{(Color online) Quantum TOF distributions $\Pi(t)$ for the coherently splitted BEC of
sodium atoms (representing the macroscopic Schr\"{o}dinger-cat) falling freely under gravity
are plotted for varying wave packet width $\sigma_0$. In each curve, time (in $sec$) is plotted along
horizontal direction and the coherent TOF distribution $\Pi(t)$ of BEC
Schr\"{o}dinger-cat is plotted (in ${sec}^{-1}$) along vertical axis. The detector is located at a
distance $z=H=-1$~cm and $d=50~\mu$m.}
\label{BECsf}
\end{figure*}

\noindent
From Fig.~\ref{Mass1}, we see that the interference pattern in quantum TOF signal gradually 
disappears as one increases the mass $m$ of the atoms. Figure \ref{BECsf} shows the quantum TOF distribution
for different values of wave packet width from $\sigma_0=1~\mu$m to $\sigma_0=6~\mu$m. It is clear
from Fig.\ref{BECsf} that the number of fringe and the contrast of interference pattern in quantum 
TOF distribution decreases as one increases the value of the initial widths ($\sigma_0$) of the wave 
packets. Nevertheless, it is possible to see the interference for a larger value of $\sigma_0$.
For example, if one chooses $\sigma_0=10~\mu$m, then to observe good interference pattern (with good 
contrast and having considerable number of fringe) in $\Pi(t)$, the separation $d$ needs to be considered
in the range of $50~\mu$m to $250~\mu$m, with the detector placed at a longer distance ($H=-100$ cm) 
for a fixed mass of sodium atoms. 

\vskip 0.5cm
\noindent
The interference in $\Pi(t)$ is sensitive to the parameters $\sigma_0$ and the atomic mass $m$, the detector 
location $H$ and the separation $d$. We repeat here that the interference in $\Pi(t)$ arises mainly because 
of the temporal overlap $P_{12}(z=H,t)$ (\ref{overlap1}) and the oscillatory factor $\delta$ (\ref{delta}). 
To increase the temporal overlap $P_{12}(z=H,t)$, one has to find the condition under which the spreading 
of the wave packet increases: small $\sigma_0$, lighter mass atoms, distant detector location (large $H$) 
will be helpful in this regard to enhance this effect. The oscillatory factor $\delta$ can be increased 
either by reducing the value of $\sigma_0$, or by increasing the parameters $d$ and $H$. Actually, when one 
considers higher values of the parameter $\sigma_0$, then the temporal overlap $P_{12}(z=H,t)$ and the oscillatory 
factor $\delta$ both decrease. This is because for larger values of the parameters $\sigma_0$ (or mass $m$), 
the spreading effect (\ref{spread}) and hence the temporal overlap $P_{12}(z=H,t)$ becomes small. As a result, 
the wave packets try to localize (in time as well as in space) more strongly causing the interference effect 
to be small. Also, for higher values of $\sigma_0$, the oscillatory factor $\delta$ will be too small due to the 
presence of ${\sigma^4_0}$ in the denominator of $\delta$ (\ref{delta}). Then one has to allow the BEC to 
travel a longer distance (by increasing $H$) to develope some temporal overlap of the wave packets, and also 
increasing $H$ helps us to increase $\delta$ (\ref{delta}). For higher values of $\sigma_0$, the parameter 
$\delta$ should also be increased by increasing the value of the separation $d$, keeping in mind that there is 
a considerable temporal overlap $P_{12}(z=H,t)$. The temporal overlap gets reduced if one increases the separation 
$d$ too much. So, even if there is a delicate choice of the parameters, one can observe the interference for a 
wide range of parameter values. 

\begin{figure*}
\begin{center}
\includegraphics{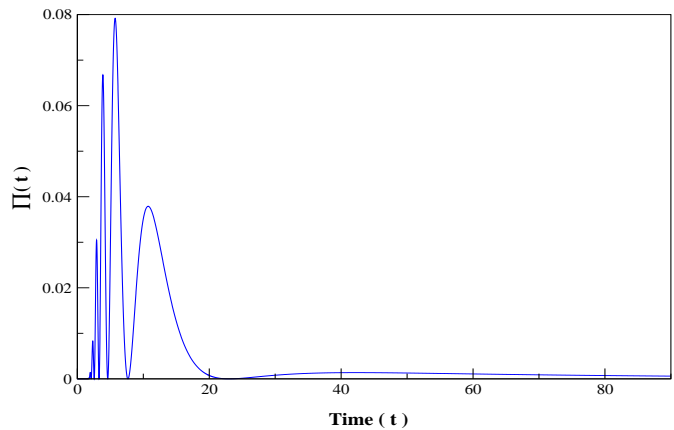}
\caption{(Color online) Quantum TOF distribution $\Pi(t)$ for the coherently splitted BEC of
sodium atoms (representing the macroscopic Schr\"{o}dinger-cat) is plotted in the absence of 
gravity ($g=0$). Time (in $sec$) is plotted along horizontal direction and the coherent TOF
distribution $\Pi(t)$ of BEC Schr\"{o}dinger-cat is plotted (in ${sec}^{-1}$) along vertical axis.
The detector is located at a distance $z=H=-1$~cm with $\sigma_0=1~\mu$m and $d=20~\mu$m.}
\label{Gravity0}
\end{center}
\end{figure*}

\vskip 0.5cm
\noindent
It is significant to mention here that gravity plays an important role in our setup to observe the interference in 
the TOF distribution. In Fig.\ref{Gravity0} we plot the time distribution $\Pi(t)$ for $g=0$ (no gravity) and compare 
it with Fig.(\ref{BECdf}d) where we plot $\Pi(t)$ in the presence of gravity with the parameter values same as that of 
Fig.\ref{Gravity0}. We see that the magnitude of $\Pi(t)$ (in the absence of gravity) is roughly $10^{5}$ times smaller 
than that obtained for gravitational free fall case for $z=H=-1$ cm. The reason for this is that the magnitudes of 
$J_1(z=H,t)$ and $J_2(z=H,t)$ become very small (roughly $10^{5}$ times) in the absence of gravity. This magnitude 
becomes $10^{6}$ times smaller if we consider the detector location at $z=-10$ cm. Actually, in the absence of 
gravity, there will be free particle motion and expansion of the wave packets in every direction. So, if one tries 
to observe the interference in the quantum TOF distribution in the absence of gravity, the intensity of that 
interference pattern will be too faint to be observed as only a small fraction of the condensate atoms will arrive 
at the detector. Hence, in our setup, gravity plays an important role which helps to pull down the condensate towards 
the detection plane.

\section{Summary and Conclusion}\label{sec:Conclusion}

To summarize, in this work we propose a scheme to experimentally observe matter-wave interference 
in the time domain, specifically in the TOF  (arrival-time) distribution using atomic BEC. This 
experimentally testable scheme has the potential to empirically resolve ambiguities inherent in
the theoretical formulations of the quantum arrival time distribution. Here we use the probability 
current density approach to calculate the quantum TOF distributions for atomic BEC 
Schr\"{o}dinger cat represented by superposition of macroscopically separated wave packets in space. 
Our definition of the quantum TOF distribution in terms of the modulus of the probability current 
density is particularly motivated from the equation of continuity, and other physical considerations 
discussed in the literature \cite{timereport,current,rotator,finkel,bohmarrtim,ali1,ali2,class,weq}. 
This approach also provides a proper 
classical limit, as the interference and hence the coherence in the quantum TOF signal disappears 
in the large-mass limit. We repeat that there is no classical analogue of this TOF distribution 
$\Pi(t)$ and this is purely a quantum distribution where we see the matter-wave interference in 
the quantum TOF signal. Hence, it will be interesting to see if our prediction of interference in 
time domain (TOF distribution) can be verified in actual experiments using modern interferometry 
techniques and sophisticated TOF methods.

\vskip 0.2cm
{\bf Acknowledgments}
\vskip 0.2cm
\noindent
We would like to acknowledge support from the National Science
Council, Taiwan, under Grant No. 97-2112-M-002-012-MY3, 
support from the Frontier and Innovative Research Program 
of the National Taiwan University under Grants No. 97R0066-65 and 
No. 97R0066-67, and support from the focus group program of the National Center for Theoretical Sciences, 
Taiwan. H.S.G. is grateful to the National Center for High-performance Computing, Taiwan, 
for computer time and facilities.

\end{document}